\documentclass[showpacs,preprint,amsmath,superscriptaddress,prl]{revtex4}  
\usepackage{epsfig}
\usepackage{epstopdf} 
\usepackage{graphicx}
\usepackage{color}

\newcommand{\elabel}[1]{\label{eq:#1}}

\newcommand{\flabel}[1]{\label{fig:#1}}
\newcommand{\fref}[1]{Fig.~\ref{fig:#1}}
\newcommand{\Fref}[1]{Figure ~\ref{fig:#1}}

\begin{document} 
\title{
Data-Driven Prediction of Thresholded Time
Series of Rainfall and SOC models \\
} 
\author{Anna Deluca}
\email{adelucasilberberg@gmail.com} 
\affiliation{
Max-Planck-Institut f\"ur Physik Komplexer Systeme, N\"othnitzer Strasse 38, D-01187 
Dresden, Germany}
\author{Nicholas R. Moloney}
\affiliation{%
London Mathematical Laboratory, 14 Buckingham Street, WC2N 6DF London, UK
}
\author{\'Alvaro Corral}
\email{acorral@crm.es} 
\affiliation{Centre de Recerca Matem\`atica,
Edifici C, Campus Bellaterra,
E-08193 Barcelona, Spain.
} 
\affiliation{Departament de Matem\`atiques,
Facultat de Ci\`encies,
Universitat Aut\`onoma de Barcelona,
E-08193 Barcelona, Spain}
\begin{abstract} 
We study the occurrence of events, subject to threshold, in a
representative SOC sandpile model and in high-resolution rainfall
data.  The predictability in both systems is analyzed by means of a
decision variable sensitive to event clustering, and the quality of
the predictions is evaluated by the receiver operating characteristics
(ROC) method.  In the case of the SOC sandpile model, the scaling of
quiet-time distributions with increasing threshold leads to increased
predictability of extreme events.  A scaling theory allows us to
understand all the details of the prediction procedure and to
extrapolate the shape of the ROC curves for the most extreme events.
For rainfall data, the quiet-time distributions do not scale for high
thresholds, which means that the corresponding ROC curves cannot be
straightforwardly related to those for lower thresholds.
\end{abstract} 

\maketitle

\section{Introduction}

Many atmospheric processes related to precipitation give rise to
structures and correlations across long ranges in space and time,
which are the result of the coupling between several non-linear
mechanisms with different spatial and temporal characteristic scales
\cite{Lovejoy1982area,vattayharnos,Yano2003,BodenschatzETAL2010}.
Despite the diversity of individual rain events, an array of
statistical measures presents strong statistical regularities
\cite{Peters_prl,
  Peters_np,Neelin_rethinking,Peters_Deluca,Deluca_npg,DelucaPuigETAL2013},
giving support to the hypothesis that atmospheric convection and
precipitation may be a real-world example of self-organized
criticality (SOC)~\citep{BTW87}.  Whereas the usual approach in
meteorology and hydrology consists of looking at the occurrence of
rain in fixed time intervals (days, months...), 
``episodic" rain events, similar to avalanches in SOC sandpile models,
can be defined by integrating the rain rate over very short
time periods. This led to the claim that rain-event sizes are
power-law (i.e., scale-free) distributed, at least in the unique site
studied
\citep{Peters_prl}, in agreement with the SOC hypothesis.
Nevertheless, a power-law distribution of this observable is not
sufficient evidence for SOC dynamics, 
as there are many alternative mechanisms that give rise to such
behavior (see, for example, Refs.~\cite{Dickman2003,Mitz}).

Further support for the SOC hypothesis was given by Peters and Neelin
\cite{Peters_np} who found, for rain data over the tropical oceans,
(i) a relation between satellite estimates of rain rate and
water vapor compatible with a phase transition, with large parts of
the troposphere in a convectively active phase; and (ii) that the
system was close to the transition point most of the time.  This
constitutes genuine evidence for SOC.  These authors also related this
SOC behavior to the concept of atmospheric quasi-equilibrium
\citep{ArakawaSchubert1974}, which argues that, since driven
processes are generally slow compared to convection, the system should
typically be in a far-from equilibrium statistically stationary state,
where driving and dissipation are in balance.

Coming back to local event-size distributions, recent works have shown
that these are indicative of universality as expected in the SOC
framework ~\citep{Peters_Deluca,Deluca_npg,DelucaPuigETAL2013}.
The resulting rain-event-size distributions for several sites
distributed worldwide are well approximated by power laws of similar
exponents over broad fitting ranges, with differences only in the
large-scale cut-offs of the distributions.  These differences are
attributed to finite-size effects, pointing to distinct system
capacities in different places.

This SOC framework raises the question of the possible implications of
the critical behavior of atmospheric convection for the prediction of
rainfall, which remain unclear, due to the fact that predictability in
SOC systems is still not well understood.  In particular, prediction
studies often focus on the behavior of the most hazardous episodes,
i.e. extreme events of rainfall in our case.

In a more general context, Kantz \cite{Kantz2010} classifies the
different scientific approaches to extreme events, such as
\textit{extreme value statistics} for the robust estimation of the
tails; \textit{data driven predictions}, which employ conditional
probabilities and temporal correlations; \textit{simplistic physical
  models} to investigate the mechanisms and dynamics from which
extreme events emerge; and \textit{detailed disciplinary
  investigations} of extreme events in a particular system.

Our approach in this paper combines simplistic physical models
(sandpile models) and data driven prediction in order to gain insight
into rainfall occurrence, as well as a deeper understanding of SOC
phenomena. Because a direct connection between sandpile-like models
and rainfall has not yet been established, our purpose is not the
modeling of the latter by the former, but rather the comparison of the
dynamics of the two systems.

In the next section, we introduce the rain database and the SOC
sandpile model used for comparison, and we define the extreme events
of interest and present distributions of the quiet times that separate
them. In Section III, we explain the prediction procedure, which makes
use of the hazard function, and evaluate {the predictability} via the receiver operating
characteristics (ROC) method.  Section IV explores the prediction
procedure analytically.



\section{Rainfall data and SOC models} 

We analyze high-resolution local rain intensities across different
climates from the Atmospheric Radiation Measurement (ARM) database,
and simulated data from the Manna sandpile model. The rain data
consists of point-location measurements from sites around the world,
with one minute temporal resolution spanning about 8 months to 7
years, depending on the site. For more details, see
Ref. \cite{Peters_Deluca,DelucaPuigETAL2013}.
The signal directly measured is the rain rate, giving the
``instantaneous'' depth of precipitation at a point location every
minute, with a resolution of 0.001 mm/hour.  In fact, anything below
0.2 mm/hour should be considered as zero, because it is not possible
to distinguish rainfall from other phenomena such as mist.  A rain
``event'' is defined as a sequence of rates exceeding a threshold
$r_c$. To be precise, the event starts when the threshold is exceeded
for the first time and ends when the rate subsequently falls below
threshold. Previous works on rainfall considered only the minimum
possible threshold. However, this is not a fundamental physical
parameter but rather an unavoidable limitation of the observational
procedure. In this paper we are interested in exploring a range of
thresholds, starting at the minimum reliable value of 0.2 mm/hour and
working up to extreme thresholds.


As a prototypical SOC system we investigate the Abelian Manna
sandpile model~\citep{Dhar99} in two dimensions.
%
%
The model is defined on a
two-dimensional square lattice of size $L^2$ with open boundaries. Each
site $i = 1,\ldots,L^2$ contains a discrete number of particles
$z_{i}$.  The rules of the dynamics are:
\begin{enumerate}
\item Driving. A particle is added to a randomly chosen site $i$,
  $z_{i} \to z_{i}+1$.
\item Toppling. If $z_{i} > 1$ at any site $i$, the site is relaxed,
  $z_{i} \to z_{i}-2$, and {the} two particles are distributed among
  randomly and independently chosen nearest neighbours (nn),
  $z_{\text{nn}} \to z_{\text{nn}}+1$ (with possibly the same site
  chosen twice).  Multiple topplings are performed in parallel. This
  rule is iterated (if necessary) and each update defines one step in
  the ``fast'' avalanche time scale.
\item Dissipation. Particles that are distributed from edge sites
  beyond the boundaries of the lattice are removed from the system.
\end{enumerate}
{In the following, we drive the system `infinitely slowly'
  (according to the usual protocol) by
  adding a new particle to the system only if all sites are relaxed.} 
{This {effects} a time-scale separation between driving and toppling. 
(For completeness, 
we also check the robustness of our results when this
time-scale separation is broken, as in Refs. \cite{Corral_Paczuski,Paczuski_btw}.)}
Once the system has reached the statistically stationary state (i.e.,
the number of particles in the pile stabilizes, on average), the
toppling activity {$n$, which counts the number of
toppling sites at each avalanche time step}, is recorded. 
As with rain data, we make use of an
activity threshold $n_c$ and define events as consecutive sequences of
$n>n_c$.

A key observable in our analysis (both for the rain data and the
sandpile model) is the quiet time, $\tau$, defined as the time the
{rate or} activity signal spends below threshold. For the
Bak-Tang-Wiesenfeld (BTW) model and other SOC models, it is well known
that, if the system is slowly driven {and the quiet time is
  measured in the slow time scale}, the quiet-time distribution is
approximately exponential when $n_c=0$, and therefore the 
{instantaneous} avalanches occur in the manner of a Poisson
process~\cite{Garber_pre}, {which, by definition, has no
  memory}. Paczuski et al.~\cite{Paczuski_btw} showed that this
continues to hold even when the time-scale separation is broken (and
one measures time on the avalanche time scale), provided $n_c=0$.

At first sight, this observation appears to rule out the application
of SOC models to real-world systems, such as solar flares or
earthquakes, for which the quiet-time distributions are not
exponential~\cite{Boffetta,Corral_prl.2004,Corral_comment}.  However, as shown
in Ref.~\cite{Paczuski_btw} for the BTW model, when a non-zero
threshold is applied ($n_c>0$), the quiet-time probability density
function (PDF) is a decaying power law with exponent $\beta \approx
1.67$, independent of system size. This non-exponential PDF reflects
the existence of clustering between events, in agreement with many
natural phenomena. In the real world, thresholds are often imposed,
e.g. because of the limited resolution of measuring devices, and
therefore it is more realistic to consider non-zero thresholds for
practical purposes.


Our simulations lead to very similar results for the Manna model 
{with infinitely slow driving.
Due to the time-scale separation, the quiet time is only defined 
inside the original, zero-threshold avalanches,
and measured in the fast time scale, as
in model A of Ref. \cite{Paczuski_btw}. 
\Fref{BTW2d_twait}(a)} shows, {for a system of size $L=1024$}, quiet-time PDFs $P_q(\tau)$,
{where $q$ denotes the thresholds as quantiles of the activity distribution.} 
For any $n_c>0$ ({i.e., $q>0$}) the PDFs are clearly compatible
with a power law with negative exponent $\beta=1.67$ up until some
fast-decaying cut-off. Although the cut-off function moves out with
increasing threshold, a convincing data collapse onto a
threshold-independent scaling function is possible, as shown in
\fref{BTW2d_twait}(b) (see also \cite{Peters_Deluca}). 
{Similar behavior is observed when time-scale separation is broken by
  adding a particle to the system every 100 avalanche steps (whether
  or not all sites are relaxed). However, the scaling is not so clean:
  a bump appears between the power law and the subsequent faster
  decay (not shown). This arises from the mixture of quiet times within
  avalanches (corresponding to the previous case, model A) and
  exponential quiet times between avalanches.}
 

\begin{figure}
\centering
\includegraphics*[height=0.68\textwidth,angle=270]
{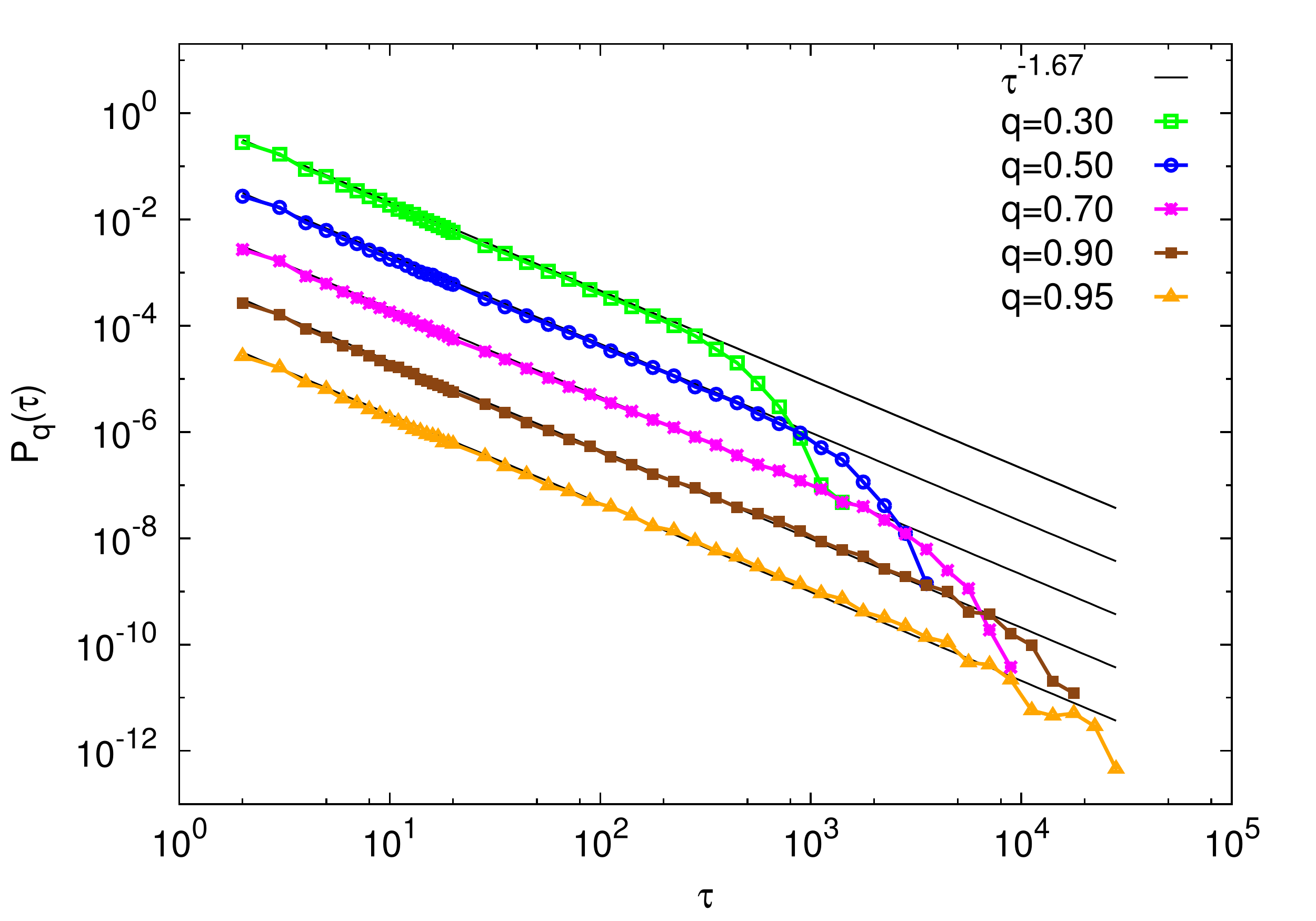}
\includegraphics*[height=0.68\textwidth,angle=270]
{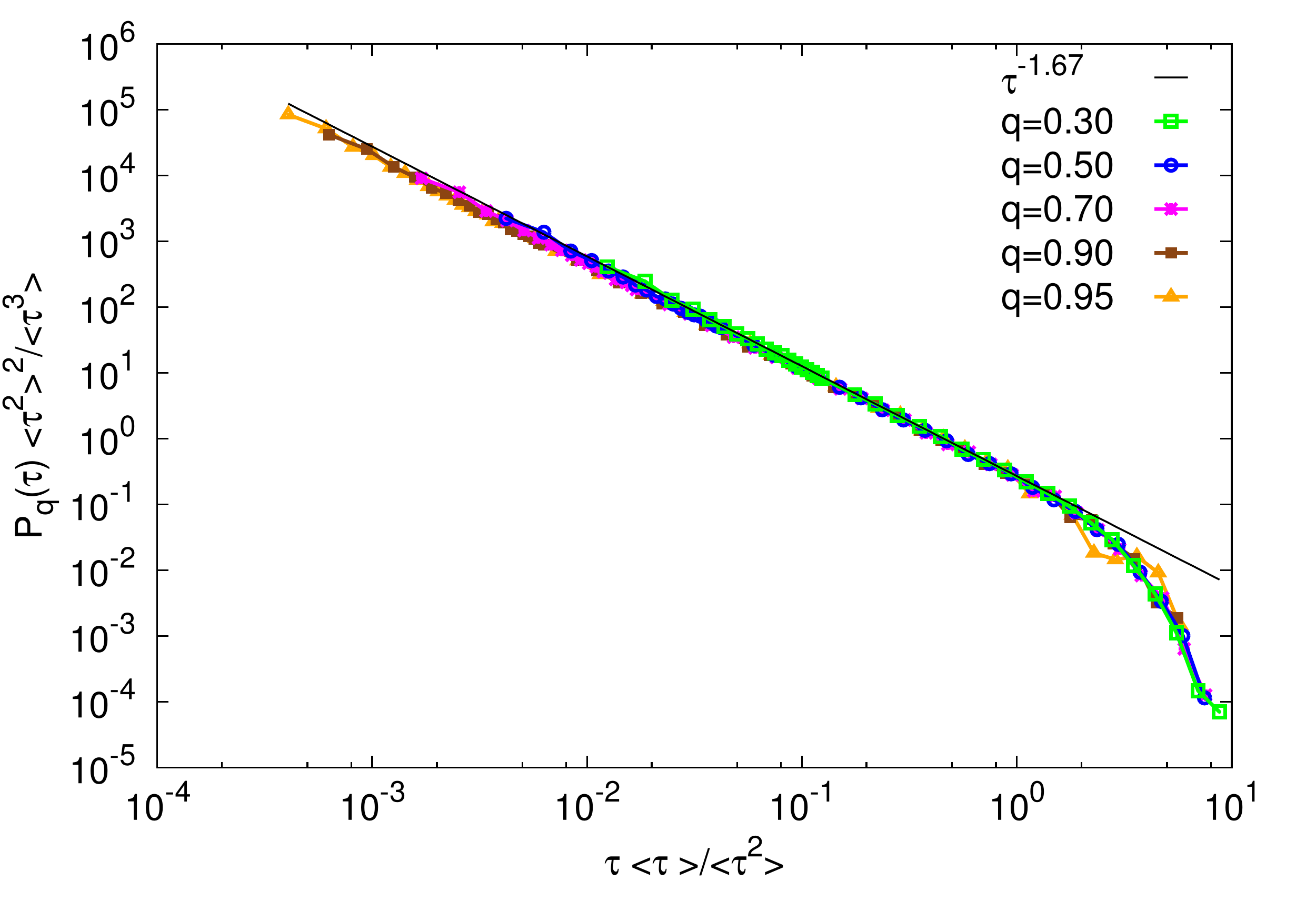}
 \caption[Quiet-time PDFs for different activity thresholds for the
   Manna model 
]  { (a) Quiet-time PDFs for
   the Manna model ($L=1024$) 
   for different activity thresholds $n_c$ characterized by their quantiles q.
Statistics are collected
   over $10^7$ particle additions. PDFs for different $q$ are
   displaced vertically for clarity. The solid black lines have slope
   $-1.67$. (b) Data collapse of the suitably rescaled quiet-time
   PDFs.
}
\flabel{BTW2d_twait}
 \end{figure}


\begin{figure} 
\begin{center}
\includegraphics*[height=0.68\textwidth,angle=270]
{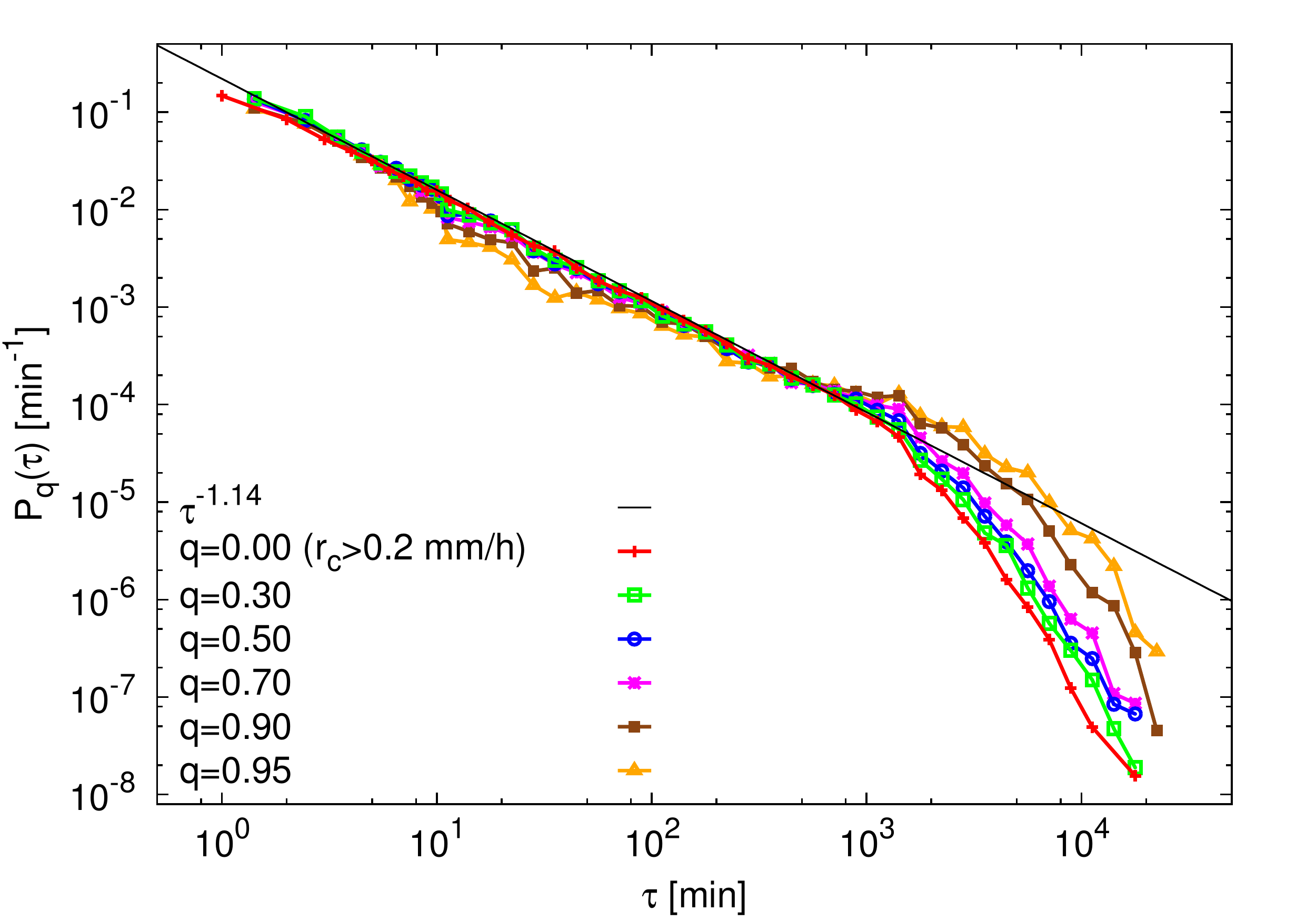}
\includegraphics*[height=0.68\textwidth,angle=270]
{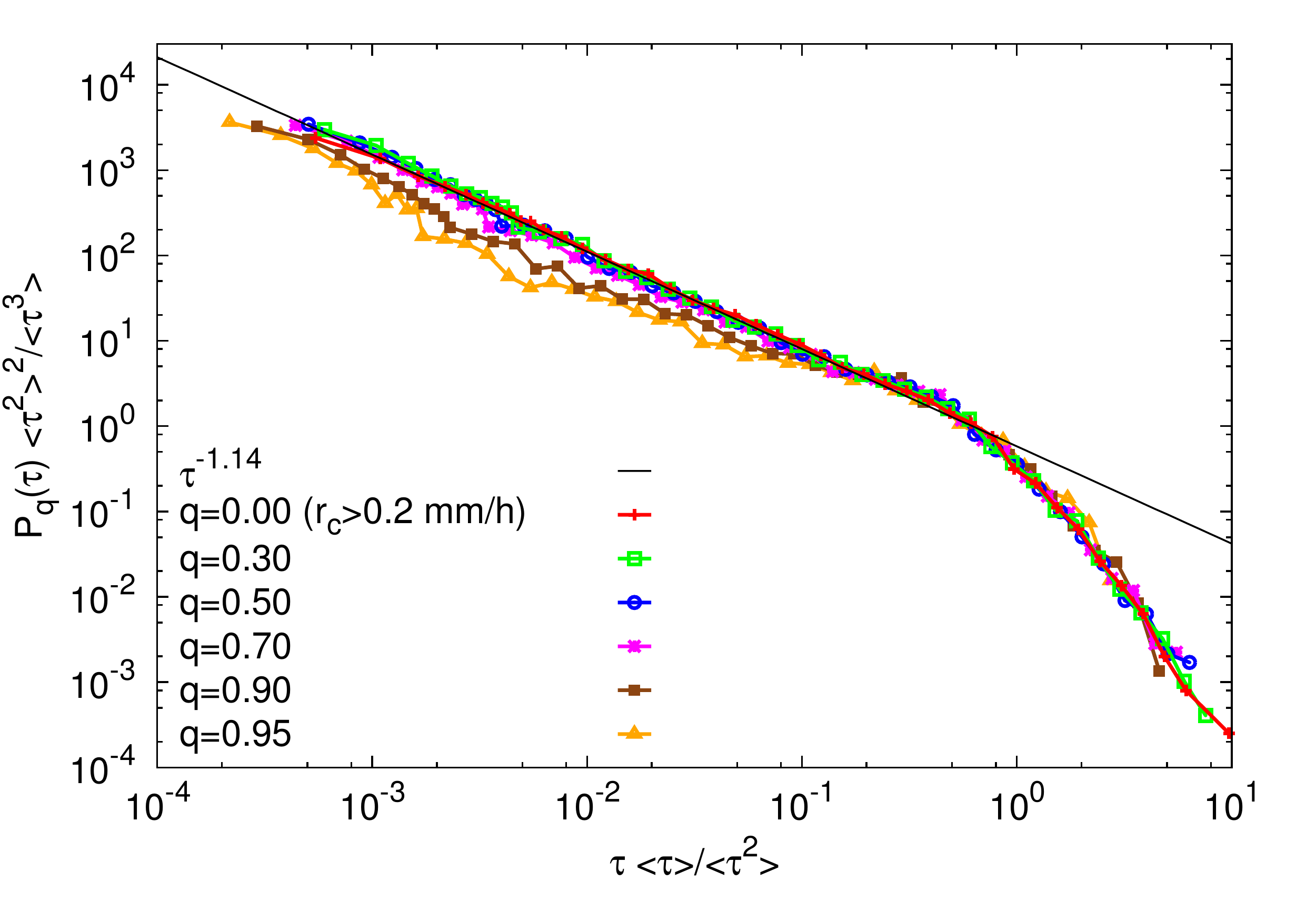}
\end{center}
\caption[Rescaled quiet time distributions for different thresholds (quantiles) on the 
{rate} for rainfall data.]
{(a) Quiet-time PDFs for different rain {rate} thresholds, as
  measured on Manus island for the period 2005/02/15 -- 2012/03/18.
A power law decay with exponent
  $\beta=1.14$ has been fitted to the $q=0.00$ data to serve as a
  visual guide. (b) Data collapse of the suitably rescaled quiet-time PDFs.
}
\flabel{Rain_data_01_twait}
\end{figure}

Turning to rainfall data, \fref{Rain_data_01_twait} plots quiet-time
PDFs $P_q(\tau)$ for different {rate} thresholds 
as measured at the ARM site on Manus island. 
{Analyses for {rate thresholds} below 0.2 mm/h are disregarded}
owing to the detection limit on measurement devices. 
Thus, even for $q=0$ an implicit threshold
exists. Contrary to the Manna model, the PDFs are approximately
independent {of} $q$ up until about $q=0.70$, 
{following a power-law decay with an exponent $\beta$ close to 
1.14
and a faster decay at the tail.
Beyond $q=0.70$}, differences become
more apparent. 
Nevertheless, a reasonable data collapse is shown in {\fref{Rain_data_01_twait}(b)}.
Other ARM sites show roughly the same behavior. 
In the next two sections, we explore how differences in the
quiet-time PDFs between the Manna model and rainfall data show up in
the predictability of rain events.

\section{Hazard function and ROC curves}

For the purposes of time series prediction, we use the hazard function
$H_q$ (which is sensitive to both the clustering and repelling of
events) as a decision variable. In comparison, the conventional
precursory pattern recognition technique requires a large amount of
data, does not capture long-term {clustering}, and has been found to perform
worse in a similar analysis {of heartbeat intervals}~\citep{BogachevETAL2009}.
The hazard function~\citep{Kalbfleisch2} gives the probability {per unit time} that
the quiet time (for events {defined by a threshold given by} quantile $q$) terminates between
$t_w$ and $t_w+dt_w$, given that it has exceeded $t_w$, as illustrated
in~\fref{fhazardfunction_sketch}. That is,
\begin{align}
H_{q}(t_w)d t_w &= 
\frac{\int_{t_w}^{t_w+d t_w}
  P_q(\tau)\mathrm{d}\tau}{\int_{t_w}^{\infty}P_q(\tau)\mathrm{d}\tau}
\\
&= \frac{P_q(t_w) d t_w} {S_q(t_w)},\elabel{fhazard}
\end{align}
where $S_q(t_w)=\int_{t_w}^{\infty} P_q(\tau)\mathrm{d}\tau$ is the
survivor function~\citep{Kalbfleisch2}, i.e.,~the probability that the
quiet time is greater than $t_w$. For future reference it is useful to
note that for exponential {quiet} times (Poisson process), the hazard
function is constant. Since the time series under consideration are
discrete, $dt_w$ corresponds to one parallel update in the Manna
model, and one minute for the rain data.
\begin{figure} 
\centering
       \includegraphics*[width=0.6\textwidth]
{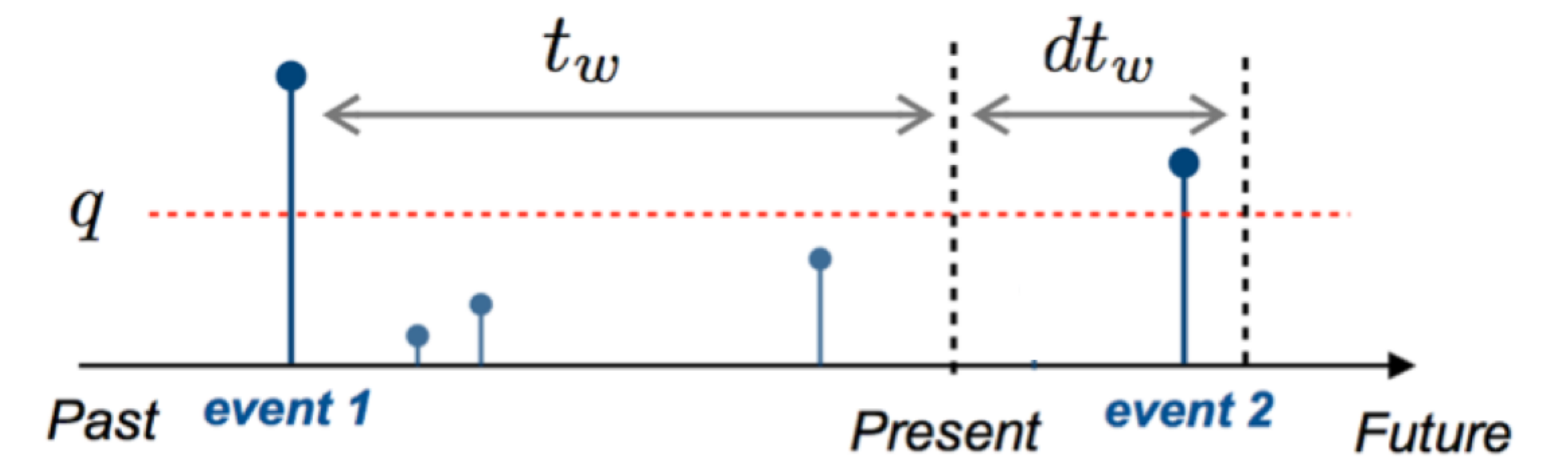}

\caption{Pictorial definition of the hazard function. Event 2 falls
  between $t_w$ and $dt_w$, having exceeded event 1 by $t_w$.
}
\flabel{fhazardfunction_sketch}
\end{figure} 
The hazard function is constructed numerically via the quiet-time PDF
and the survivor function. \Fref{hazard_fig} shows results for the
Manna model and Manus island over the complete temporal record.
\begin{figure}
\centering
      \includegraphics*[height=0.68\textwidth,angle=270]
{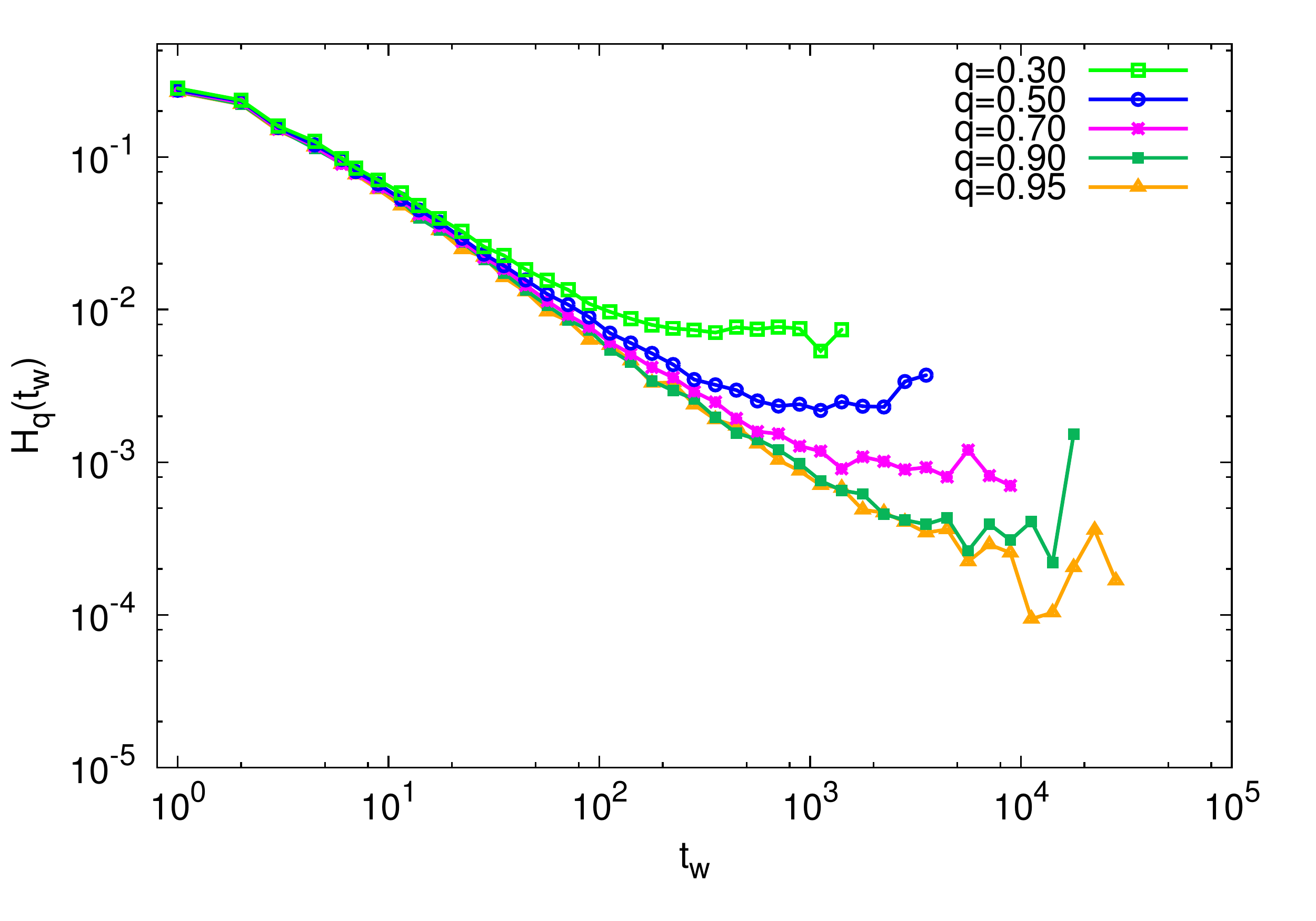}
\includegraphics*[height=0.68\textwidth,angle=270]
{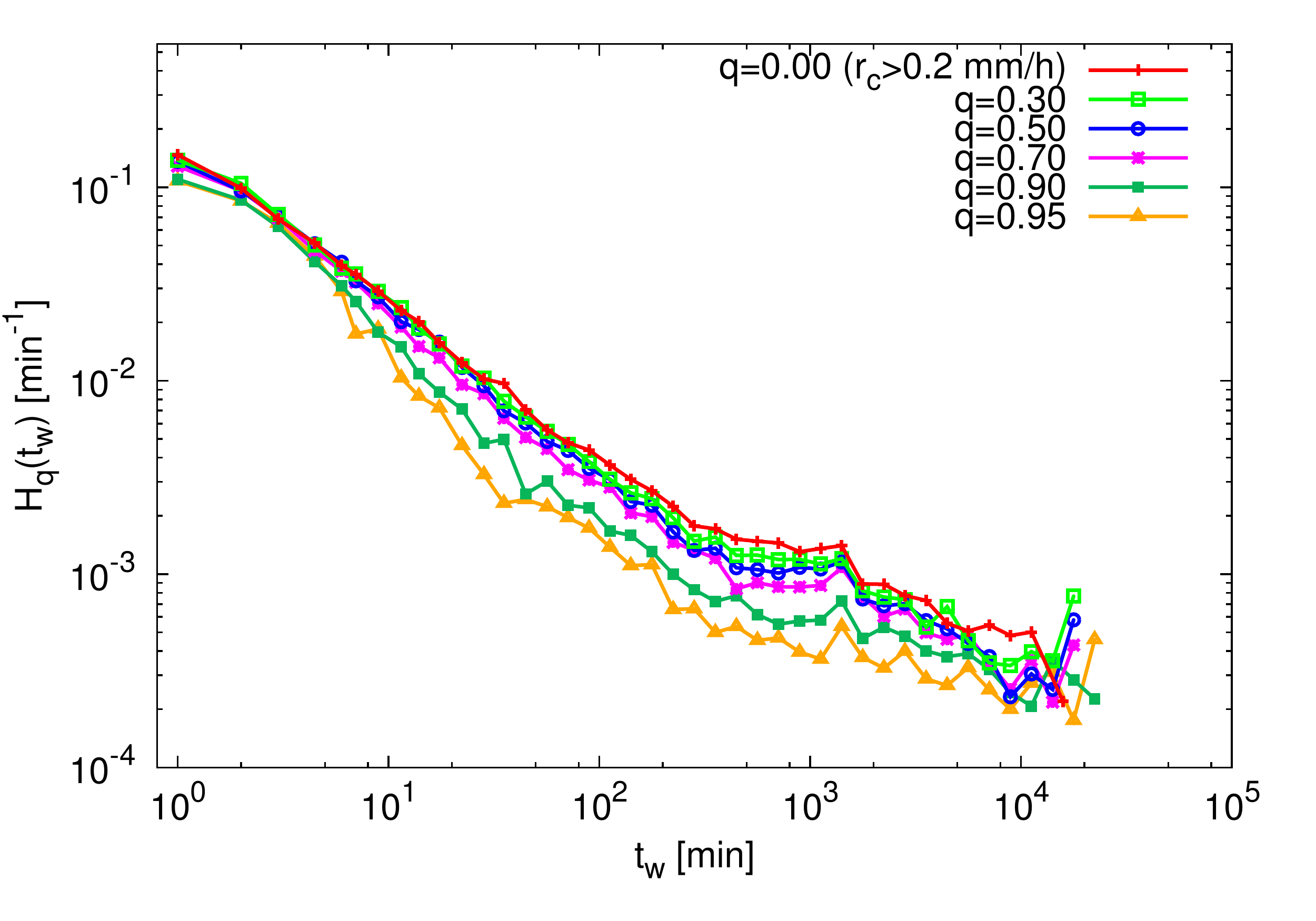}
\caption{
(a) Hazard functions for quiet times in the Manna model for different activity thresholds
(same data as in Fig. 1).
(b) Hazard functions for quiet times of Manus island rain data for different {rain rate} thresholds
(same data as in Fig. 2). 
\flabel{hazard_fig}}
\end{figure}

For the purposes of prediction we only assume access to past
information, and the hazard function is {therefore}
constructed solely from past events. We update its estimate every 100
events. Since the hazard function gives a probabilistic forecast, a
deterministic prediction is issued via a discrimination threshold.
Specifically, if in a given time step {the
  {estimated} $H_q$ exceeds the discrimination threshold,
  an alarm is raised, which is to say that an event is expected in the
  next $d t_w$}.  {If $H_q$ does not exceed the
    threshold, no alarm is raised and no event is expected.}

We evaluate the quality of the prediction using a receiver operating
characteristic (ROC)  {analysis}~\citep{Egan1975}. For any binary prediction
(alarm raised or not) there are four possible outcomes: an alarm is
raised and the event does occur (true positive, TP); an alarm is
raised and the event does not occur (false positive, FP); an alarm is
not raised and the event does occur (false negative, FN); an alarm is
not raised and the event does not occur (true negative, TN). The ROC
curve summarises this information by comparing the sensitivity (the
proportion of successfully predicted occurrences) to the specificity
(the proportion of successfully predicted non-occurrences):
\begin{equation}
\text{sensitivity}=\frac{\text{TP}}{\text{TP}+\text{FN}}, \hspace{1cm} \text{specificity}=\frac{\text{TN}}{\text{TN}+\text{FP}},\\
\end{equation}
{where TP, FN, TN, and FP refer to the number (or rate) of occurrences or non-occurrences for each case.}
Each threshold on the decision variable will give rise to a different
point on the ROC curve. For example, in the absence of a threshold an
alarm is raised every time step, such that 
FN $= 0$, yielding a sensitivity of $1$. Such a protocol will never miss an occurrence, but
will also never predict a non-occurrence, {i.e., TN $=0$}, and therefore has specificity
$0$. On the other hand, for an unsurpassable threshold an alarm is never
raised, such that 
FP $= 0$, yielding a specificity of $1$. Such
a protocol will never miss a non-occurrence, but will also never
predict an occurrence and therefore has {TP $=0$ and} sensitivity $0$. The diagonal
line in \fref{fROCdriv} that joins these two scenarios corresponds to
non-informative predictions, i.e.,~issuing alarms at some fixed rate
irrespective of the decision variable. Points above the diagonal
represent good predictions and points below poor predictions, in
comparison to totally random predictions.  The point (1,1) corresponds
to a perfect prediction.

 \begin{figure} 
\centering
\flabel{fROCMannadriv1024}
\includegraphics*[height=0.68\textwidth,angle=270]
{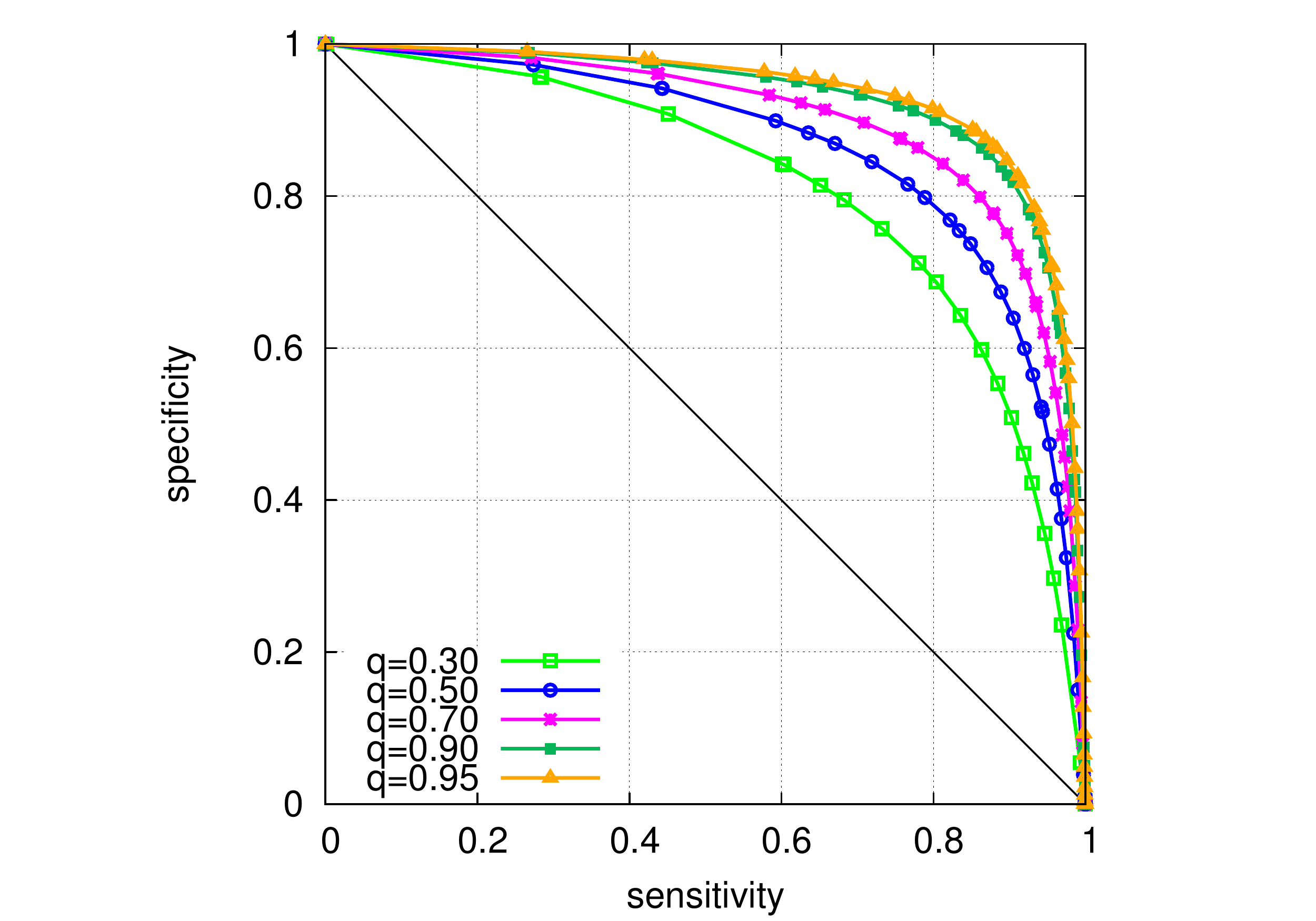}
\flabel{fROCRain_01}
\includegraphics*[height=0.68\textwidth,angle=270]
{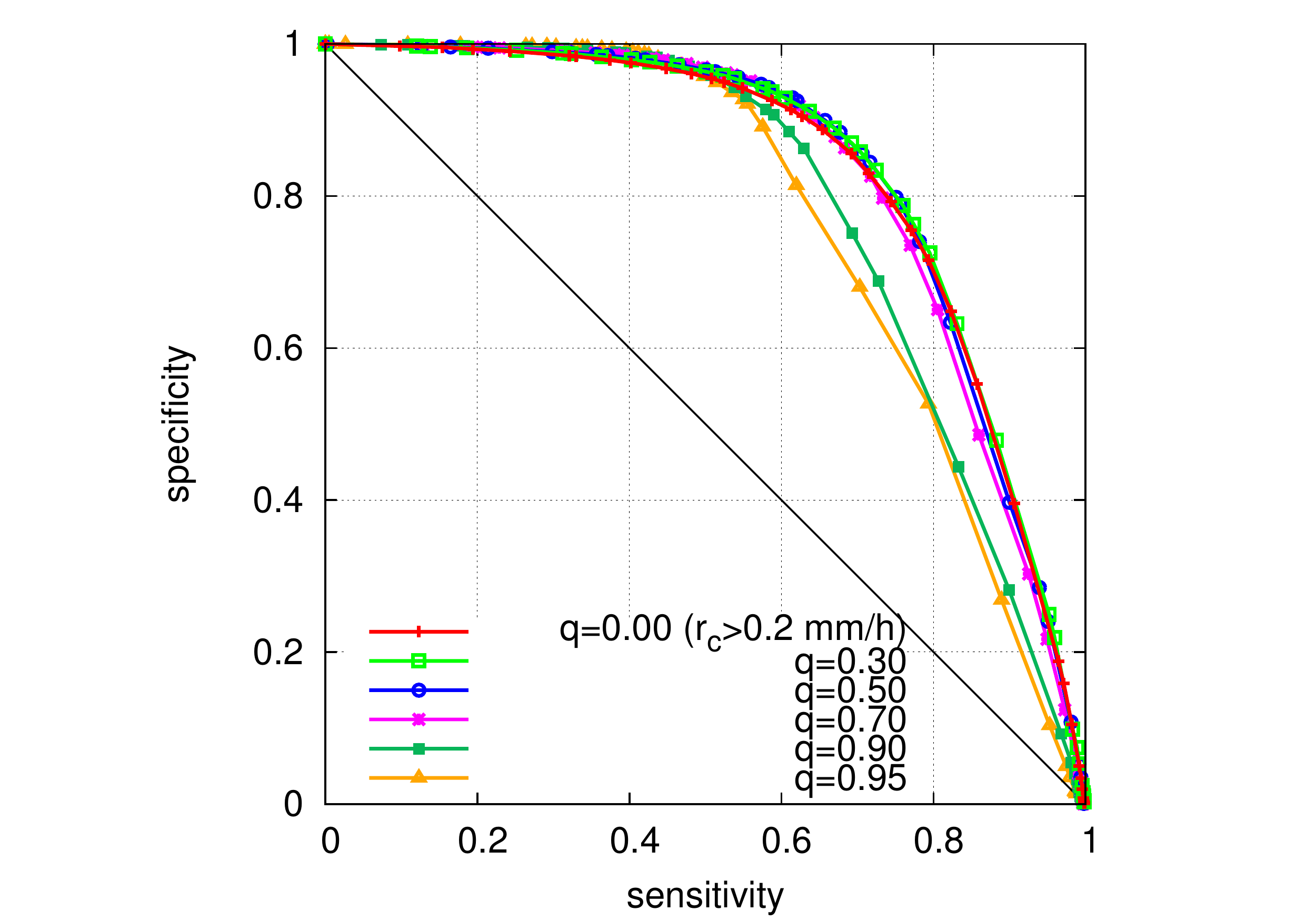}
   \caption{ROC curves for different thresholds in the {activity/rate} for 
(a) the Manna model with $L=1024$,
(b) rainfall data from Manus island.
{Same data as in Figs. 1 and 2.}
} \flabel{fROCdriv}
\end{figure}

Starting with the Manna model, the sequence of hazard functions in
\fref{hazard_fig}(a) suggests that the `memory' between events
persists for increasingly longer quiet times as the activity threshold
is increased. This is inferred from the increasing crossover times to
the constant `memoryless' portion of the hazard curve. Thus, we
anticipate that predictability improves with increasing activity
threshold. This is indeed illustrated by the ROC curves in
\fref{fROCdriv}(a), which form an ordered sequence tending towards the
corner (1,1) (perfect predictability). We will explore the clustering
that gives rise to this increased predictability in the next
section. 

For rainfall data, the picture is more complicated. 
One can observe in \fref{fROCdriv}(b)
a bundle of ROC curves which are
broadly similar up to thresholds of $q = 0.70$. For higher thresholds
we find that, for sensitivities below 0.4, predictability
(specificity) increases with threshold {in rain rate} (just as in the Manna
model), but for higher sensitivities the opposite is true. Since low
sensitivity corresponds to high thresholds in the hazard function
(i.e. only events separated by short quiet times are successfully
predicted), an increasing predictability implies that extreme events
cluster more and more over short times. Conversely, high sensitivity
corresponds to low thresholds in the hazard function (i.e. events
separated by long quiet times are also successfully predicted), and a
decreasing predictability implies that extreme events correlate less
and less over long times. A closer re-examination of the highest
quantile curves in \fref{Rain_data_01_twait}(a) bears this out: the
power-law exponent $\beta$ appears to increase in the left part of the
distribution, and the range of the exponential tail appears to
increase in the right part. These longer times could be affected by
seasonal effects which are difficult to resolve owing to the
relatively short duration of the time series (7 years). Longer time
series would be required in order to make better comparisons with
SOC-like dynamics.

\section{Analytical Treatment}

For monotonically decaying hazard functions (as is approximately the
case for the Manna model and rainfall, {see \fref{hazard_fig}}), 
raising an alarm whenever the
function exceeds a threshold value is equivalent to raising an alarm
all the while
{the elapsed time since the last event has not}
yet exceeded a `threshold' time. This threshold time, $\Delta$, is
uniquely determined by the threshold imposed on the hazard
function. Under these conditions it is possible to infer some general
behaviour for the sensitivity and specificity. For simplicity, we will
ignore any dependence on activity threshold in the following.

First, consider the sensitivity. Given an event, the subsequent event
occurs either before the threshold time, $t_w \le \Delta$, or after
the threshold time $t_w > \Delta$. In the first case, events are
successfully predicted, and therefore the total proportion of
successfully predicted events $\text{TP} \propto \int_0^\Delta
P(\tau)d\tau=1-S(\Delta)$. The remaining proportion $S(\Delta) \propto
\text{FN}$ falls to unpredicted events, i.e.,~false negatives (recall
the similarity {with} Type I errors in statistics). Thus,
\begin{equation*}
\mbox{sensitivity } = 1-S(\Delta),
\end{equation*}
which, {if we introduce $\alpha=S(\Delta)$,} 
is independent of the underlying form of the quiet-time
distribution (apart from its assumed decaying monotonicity).

Next, consider the specificity.  The specificity can also be
calculated {by instead considering
  non-events}. {Suppose an event takes place after a quiet
  time $\tau$ since the previous event. If $\tau < \Delta$ (with
  probability $1-\alpha$), the contribution to false positives is
  proportional to $\tau$, whereas the contribution to true negatives
  is zero (since the alarm will be raised all the while). If $\tau >
  \Delta$ (with probability $\alpha$), the contribution to false
  positives is proportional to $\Delta$, whereas the contribution to
  true negatives is proportional to $\tau-\Delta$ (since the alarm is
  no longer raised beyond $\Delta$).  Taking into account the
  condition $\tau \le \Delta$ in the former case, the mean rate of
  false positives is thus}
$$
\mbox{FP} \propto (1-\alpha) \langle \tau | \tau \le \Delta \rangle+ \alpha \Delta.
$$
{Following the same reasoning,
the rate of true negatives is}
$$
\mbox{TN} \propto \alpha \langle(\tau-\Delta)| \tau > \Delta \rangle.
$$
Therefore, as the proportionality factor is the same in both cases
(associated {with} the elementary time step $d t_w$),
\begin{equation}
\mbox{specificity }=\frac{\alpha \langle (\tau-\Delta)| \tau > \Delta \rangle}
{\alpha \langle (\tau-\Delta)| \tau > \Delta \rangle
+ (1-\alpha) \langle \tau | \tau \le \Delta \rangle+ \alpha \Delta}
=\frac{\alpha \langle (\tau-\Delta)| \tau > \Delta \rangle}
{\langle \tau \rangle},
\label{spec_eq}
\end{equation}
where $\langle \tau\rangle$ is the (unconditional) mean quiet time and
$\langle (\tau-\Delta)| \tau > \Delta \rangle$ is the mean residual
quiet time after an elapsed time $\Delta$.  This quantity has some
counterintuitive properties in the case of a decreasing hazard
function \cite{Davis,Corral_pre.2005}, but characterizes a random
variable in the same way as its probability distribution (if the mean
is finite \cite{Kalbfleisch2}). In contrast to the sensitivity, the
specificity {does depend on the underlying form of the
  quiet-time distribution}.

Eq. (\ref{spec_eq}) provides the relation between specificity and
sensitivity {via $\Delta$}. {For example,
  in the Poisson process the mean residual quiet time is equal to the
  mean quiet time irrespective of any elapsed time. Thus, specificity
  $=\alpha$ and so specificity $ = 1- $ sensitivity}.


It is instructive to apply the above analysis to a specific form of
the quiet-time distribution. For the Manna model, 
{this can be roughly modelled by}
a truncated gamma distribution
\begin{equation}
P(\tau) = \frac 1 {a \Gamma(\gamma,m/a)} \left(\frac a {\tau}\right)^{1-\gamma}
e^{-\tau/a},
\label{E:gamma_example}
\end{equation}
where $m \ge 0$ is the lower cut-off of the distribution. The shape
parameter $\gamma > 0$ for $m=0$ and $-\infty < \gamma < \infty$ for
$m> 0$, and the scale parameter $a>0$ (which increases with {activity}
threshold $n_c$, see \fref{BTW2d_twait}(a)). The normalizing factor
$\Gamma(\gamma,m/a)$ is the (upper) incomplete gamma function, defined
by $\Gamma(\gamma,z)=\int_z^\infty x^{\gamma-1} e^{-x} dx.$ 
Note that with this parameterization the power-law exponent
is $\beta=1-\gamma$. Thus, comparing with the results of Sec. II for
the Manna model, we have $\gamma=-0.67$. Also, exponential {quiet}
times ({as in a} Poisson process) can be recovered as a special case by
taking $\gamma=1$, $m=0$, and recalling that 
$\Gamma(1,z)=e^{-z}$.

From Eq.~\eqref{E:gamma_example}, the survivor function
\begin{equation}
S(\tau) = \int_{\tau}^\infty P(t_w) dt_w = \frac{\Gamma(\gamma,\tau/a)}{\Gamma(\gamma,m/a)},
\end{equation}
and the hazard function
\begin{equation}
H(\tau)=\frac{P(\tau)}{S(\tau)}
= \frac 1 {a \Gamma(\gamma,\tau/a)} \left(\frac a {\tau}\right)^{1-\gamma} e^{-\tau/a}.
\end{equation}
To compute the specificity, we require the mean quiet and mean
residual quiet times, which are given by
\begin{equation}
\langle \tau \rangle = a \frac{\Gamma(\gamma+1,m/a)}{\Gamma(\gamma,m/a)}
\end{equation}
and 
\begin{align}
\langle \tau - \Delta | \tau > \Delta \rangle &=
\int_m^\infty (\tau-\Delta) {P(\tau| \tau > \Delta)} d\tau=
\int_\Delta^\infty (\tau-\Delta) \frac {P(\tau)}{S(\Delta)}d\tau \\
&= \frac a {\Gamma(\gamma,\Delta/a)} \int_\Delta^\infty 
\left(\frac{\tau}{a}\right)^\gamma e^{-\tau/a} \frac {d\tau} a -\Delta
= a \frac{\Gamma(\gamma+1,\Delta/a)}{\Gamma(\gamma,\Delta/a)}-\Delta,
\end{align}
{for $\Delta > m$. This}
equation can be recast in a form more suggestive of scaling
with the help of the identity $\Gamma(\gamma+1,z)=z^\gamma e^{-z}
+\gamma \Gamma(\gamma,z)$, giving
\begin{equation}
\langle \tau - \Delta | \tau > \Delta \rangle =
a \left[\left(\frac \Delta a\right)^\gamma \frac
  {e^{-\Delta/a}}{\Gamma(\gamma,\Delta/a)} + \gamma -\frac \Delta a \right].
\end{equation}


As noted previously, high {activity} thresholds correspond to
large values of the scale parameter $a$, {while} the lower
cut-off $m$ is essentially fixed by the available time resolution,
e.g. one time step in the Manna model. Therefore, in order to explore
scaling behaviour of the sensitivity and specificity, we consider the
limit $m/a \rightarrow 0$ (with $m$ constant), in which case
\begin{equation}
\frac 1 {\Gamma(\gamma,m/a)} \simeq 
\begin{cases} 
1 / \Gamma(\gamma) & \quad \gamma > 0 \\
-\gamma (m/a)^{-\gamma} & \quad \gamma < 0,
\end{cases}
\end{equation} 
with $\Gamma(\gamma)=\Gamma(\gamma,0)$ if $\gamma > 0$,
see Ref. \cite{Abramowitz}.  
Proceeding to the quantities that appear in the ROC curves, we have
\begin{equation}
S(\Delta) \simeq 
\begin{cases}
f_+(\Delta/a) & \quad \gamma > 0 \\
a^\gamma f_-(\Delta/a) & \quad \gamma < 0, 
\end{cases}
\end{equation}
\begin{equation}
\langle \tau-\Delta | \tau > \Delta\rangle \simeq a f_2(\Delta/a)
\quad \mbox{ for all } \gamma
\end{equation}
\begin{equation}
\langle \tau \rangle \simeq 
\begin{cases}
a \gamma & \quad \gamma > 0 \\
- a^{1+\gamma} {m^{-\gamma}}{\gamma \Gamma(\gamma+1)} & \quad -1 <
\gamma < 0 \\
m\gamma /(\gamma+1) & \quad \gamma < -1,
\end{cases}
\end{equation}
where $f_+$, $f_-$, and $f_2$ are different scaling
functions. Collecting these results, we have
\begin{equation}
\mbox{specificity} = 
\begin{cases}
f_3(\Delta/a) & \quad \gamma > -1 \\
a^{\gamma+1}f_3(\Delta/a) & \quad \gamma < -1,
\end{cases}
\end{equation}
where $f_3$ is another scaling function. The sensitivity $ = 1 -
S(\Delta)$, meanwhile, scales differently for $\gamma >0 $ and $\gamma < 0$.
Thus, we finally obtain
\begin{equation}
1-\mbox{sensitivity} = 
\begin{cases}
f_4(\mbox{specificity}) & \quad \gamma > 0 \\
a^\gamma f_4(\mbox{specificity}) & \quad -1 < \gamma < 0 \\
a^\gamma f_4(\mbox{specificity}/a^{\gamma+1}) & \quad \gamma < -1,
\end{cases}
\end{equation}
where $f_4$ is another scaling function. An important consequence of
these results is that for $\gamma > 0$ predictability is independent
of $a$, whereas for {$-1 <\gamma < 0$} sensitivity increases with $a$ for a
fixed specificity. In this sense, extreme events are more predictable,
{as in Ref. \cite{Garber_pre}, but for different reasons}.
In fact, the sensitivity tends to one
as $a \rightarrow \infty$ for any non-zero specificity.
This is in agreement with our findings for the Manna model.

A concrete validation of this analysis can be achieved 
by plotting, 
$$
\mbox{specificity}
\mbox { versus }
\frac{1-\mbox{sensitivity}}{\langle \tau^2 \rangle^\gamma / \langle \tau \rangle^\gamma },
$$
which makes use of the fact that the scale parameter $a$ is proportional to 
$\langle \tau^2 \rangle / \langle \tau \rangle$  {when $ -1<\gamma <0$}, 
see Refs. \cite{Peters_Deluca,Corral_csf}.
{{Noting} that $a^\gamma \propto \langle \tau \rangle^2 / \langle \tau^2 \rangle$,
a non-parametric version of the scaling law for the ROC curve reads
$$
\mbox{specificity}
\mbox { versus }
\frac{1-\mbox{sensitivity}}{\langle \tau \rangle^2 / \langle \tau^2 \rangle },
$$
{which is also valid for {when $ -1<\gamma <0$}} \cite{Corral_csf}, and displayed in \fref{fROCMannacollapse} for the Manna model.
The scaling is reasonable but not perfect. {For a better
approximation the finiteness of the time resolution,
i.e., the fact that $m<0$ should be take into account
in the calculation of the specificity, Eq. (\ref{spec_eq}}). 
If $m$ is not negligible, then the FP rate is reduced by a term proportional to $m$,
which leads to the replacement of $\langle \tau \rangle$ by $\langle \tau \rangle-m$
in the denominator of Eq. \ref{spec_eq}. This means that all the
previous equations remain valid if the specificity is multiplied
by a factor $1-m /\langle \tau \rangle$. {Thus, the scaling of the ROC
curve is generalized to}
$$
\mbox{specificity}\times \left(1-\frac m {\langle \tau \rangle}\right)
\mbox { versus }
\frac{1-\mbox{sensitivity}}{\langle \tau \rangle^2 / \langle \tau^2 \rangle }.
$$
}
{Although this scaling theory works well for the Manna model, 
the collapse of the ROC curves for rainfall is {poor (not shown),
  because the quiet-time distributions do not scale for high rain-rate thresholds}.}

\begin{figure} 
\centering
\includegraphics*[height=0.68\textwidth,angle=270]
{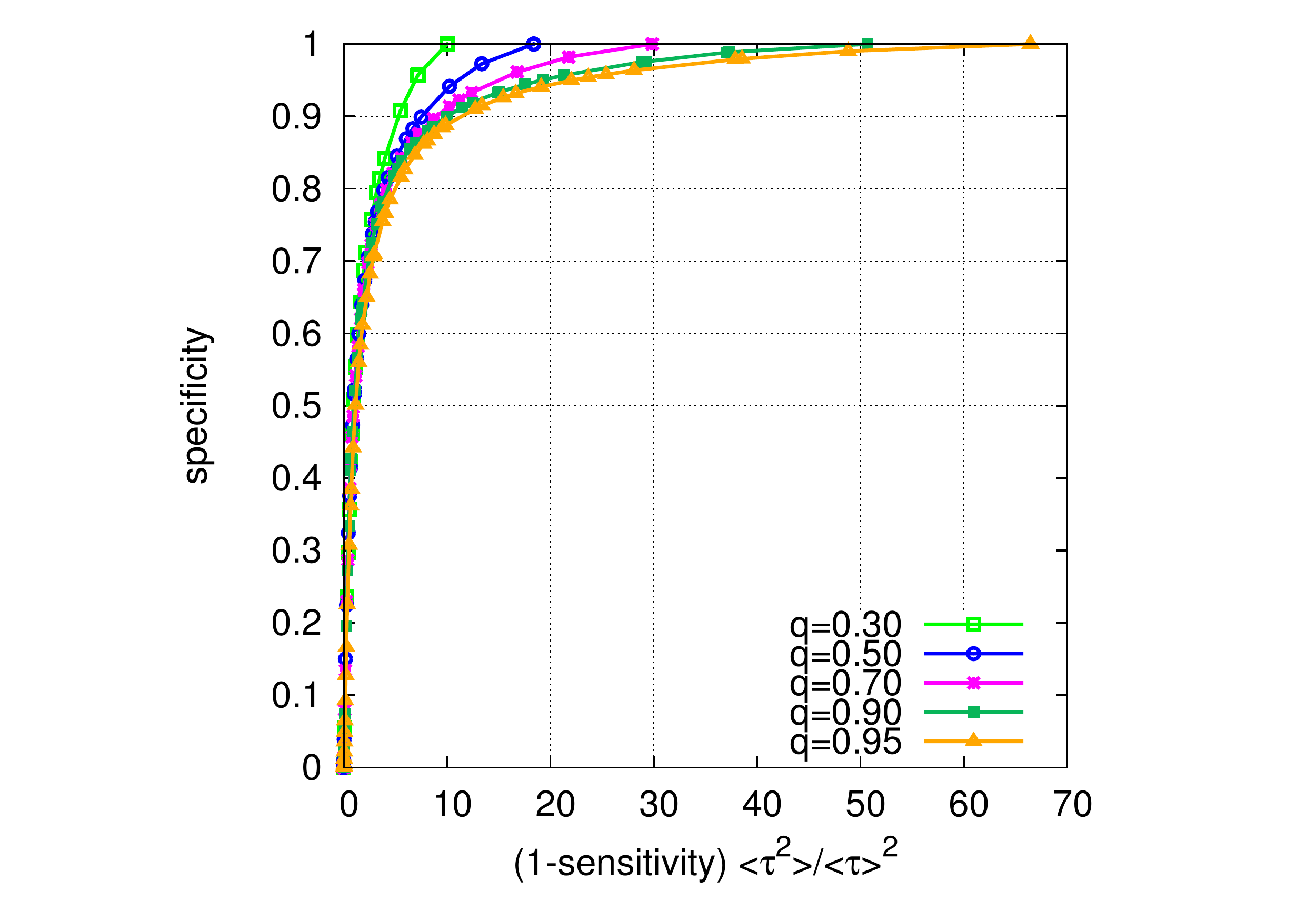}
\caption{Data collapse after rescaling the ROC curves for different activity thresholds 
in the Manna model with $L=1024$.
{Same data as in Fig. 1.}}
\flabel{fROCMannacollapse}
\end{figure}

{As a final comment,}
although we have carried out this analysis for a truncated gamma
distribution, we expect the {previous scaling results} to hold for any quiet-time
distribution of the form
$$
P(\tau) = \frac 1 {\tau^{1-\gamma}} f_0(\tau/a),
$$
where $f_0$ is a scaling function taking a constant value
for small arguments and decaying fast enough for large
arguments. The normalization constant hidden in $f_0$ may also depend
on $a$.

\section{Conclusions}

In summary, we have studied the effect of thresholding in a
representative SOC model and on actual rainfall data. The
predictability of events is studied by means of a decision variable
sensitive to the tendency of the events to cluster or repulse, and the
quality of the predictions is evaluated by the {ROC}
method. {Thresholds have been applied to the rate or
  activity, which, in the context of SOC models, corresponds to
  observing the process on the fast (avalanche) time scale}.  In this
case, the relative weight of the exponential tail decreases as the
threshold increases, leading to higher predictability.  For rainfall
data, {however}, the {change in the ROC curves
  with threshold is less clear, since the quiet-time distributions do
  not seem to scale for high thresholds}.  A scaling theory developed
for the Manna model, valid for any system {in} which the
quiet-time distribution scales with threshold, {helps us}
understand all the details of the prediction procedure.  The
philosophy of our paper is similar to that of
Ref. \cite{Hallerberg_pre08}, but note that {in that work}
the prediction scheme is based on precursory structures rather than
the hazard function.

{{Clearly}, our prediction method {works
    best} for renewal processes (point processes in which the quiet
  times are independent of each other). 
Analogies with other natural
  hazards \cite{Corral_tectono} suggest that the renewal 
process is just a first approximation. {Extensions to
    our approach would therefore include previous history, such that
    the hazard function is a function of previous quiet
    times. 
Nevertheless, our analytical approach is valid for any point process 
to which our prediction procedure is applied (not only for renewal processes).
Further refinements would include knowledge of the rain
    rate or toppling activity below threshold. These extensions are
    left for future research.}}

\section{Acknowledgements} 

Rain data was obtained from the Atmospheric Radiation Measurement Program sponsored
by the US Department of Energy, Office of Science, Office of Biological and Environmental
Research, Environmental Sciences Division. 
The authors were introduced in rain research by Ole Peters.
Research expenses were founded by
projects FIS2009-09508, from the disappeared Spanish MICINN, FIS2012-31324, from Spanish
MINECO, and 2009SGR-164 and
2014SGR-1307, from AGAUR.
A.D. is grateful for the hospitality of the Max Planck Institute for the Physics of Complex Systems
and the Universidade Federal de Minas Gerais.

%







\end{document}